\documentstyle[multicol,aps,prl]{revtex}

\begin{document}
\widetext

\draft
\title{Statistics of tumbling of a single polymer molecule in shear flow.}
\author{Sergiy Gerashchenko and Victor Steinberg}

\address{Department of Physics of Complex Systems,
         Weizmann Institute of Science, Rehovot, 76100 Israel}

\date{\today}
\maketitle
\begin{abstract}
We present experimental results on statistics of polymer
orientation angles relatively to shear plane and tumbling times in
shear flow with thermal noise. Strong deviation of probability
distribution functions (PDF) of these parameters from Gaussian was
observed and a good accord with theory was found. The scaling
relations of PDF widths for both angles as a function of the
control parameter $Wi$ are verified and compared with numerics. An
universal exponential PDF tail for the tumbling times and its
predicted scaling with $Wi$ are also tested experimentally against
numerics.

\end{abstract}

\pacs{PACS numbers: 23.23.+x, 56.65.Dy}

\begin{multicols}{2}

\narrowtext

Dynamics and statistics of a single polymer molecule in stationary
as well as in random flows have recently attracted attention of
both experimentalists\cite{chu1,chu2,chu3,sergiy} and theorists
 \cite{lebed,chertkov,hur,wang,bruno,celani}. Stretching dynamics and statistics
 and coil-stretch transition in these flows were investigated in
 detail. Besides, another remarkable effect was first observed in a shear flow\cite{chu2},
 namely large fluctuations in a polymer elongation due to
end-to-end aperiodic tumbling (see Fig.1D).

Recently rather extensive theoretical\cite{chertkov1,turitsyn} and
numerical\cite{puliafito,celani1}
  efforts were conducted with the goal to understand the statistics of the angular
 orientation of a polymer molecule and of the tumbling time in a random velocity field with a mean shear.
Shear flow with a thermal noise was considered there as a
particular case. A role of thermal noise on a solid rod tumbling
in a shear flow was first considered in Ref.\cite{hinch}. Then the
role of the Brownian fluctuations in the polymer dynamics and
statistics was studied in numerical simulations\cite{hur}, where,
however, only the power spectral density and the statistics of
polymer extension in a shear flow were investigated. An average
polymer extension and angular orientation were theoretically
considered also in Ref.\cite{wang}, where the results of
calculations were compared with light scattering
measurements\cite{link,muller}.

In this Letter we concentrate on statistics of angular orientation
and tumbling and scaling relations of their characteristics for a
single DNA molecule in a shear flow, when thermal fluctuations are
the main cause for tumbling.

It is well known that a probability distribution function (PDF) of
the end-to-end vector $\vec{R}$ for a polymer described by a
dumbbell model with a linear relaxation in a simple shear  is
Gaussian\cite{hur}. Nevertheless, PDFs of the polymer extension,
$|\vec{R}|\equiv R$, as well as polymer angular orientation are
strongly non-Gaussian due to anisotropy introduced by the shear. A
functional form of the PDF of the polymer extension in a shear
flow was first identified experimentally\cite{chu2} and then
explained theoretically and numerically\cite{hur}.

The main result of the recent theory\cite{chertkov1,turitsyn} is
the prediction of the intermittent (non-Gaussian) statistics of
polymer angular orientation in a shear flow. Moreover, it was also
shown that intermediate asymptotic of the angular polymer
statistics bears some universal features independent of the nature
of polymer random excitation\cite{chertkov1,puliafito,celani1}.

The universal features of the orientation statistics are related
to a deterministic nature of the angle evolution at large angles,
where a random pumping can be neglected. Then the
theory\cite{chertkov1,turitsyn} predicts the following results on
the angular distributions and tumbling time statistics: (i) at
$|\phi|\gg \Delta\phi\sim Wi^{-1}$ one gets $P(\phi)\propto
(\sin\phi)^{-2}$, where $\Delta\phi$ is the width of $P(\phi)$;
(ii) at $1\gg|\theta|\gg\Delta\phi$ the PDF tail for $\theta$ is
determined as $P(\theta)\propto |\theta|^{-2}$; both scalings
follow from the deterministic process; (iii) the tail of PDF of
tumbling time between two consequent flips, $\tau_{\phi}$, is
exponential at $\tau_{\phi}\gg\tau_t$, where $\tau_t$ is the
characteristic tumbling time that is proportional to $\tau_{rel}$.
Thus, the major theoretical finding is the intermittency
phenomenon in the statistics of a single polymer tumbling that
results in either algebraic or exponential tails of the
distribution of polymer angular orientations and tumbling times,
respectively.

Another theoretical result verified by extensive numerical
simulations based on linear dumbbell as well as FENE models
describes the scaling relations for the width of the angular
distributions and the characteristic tumbling time with
$Wi$\cite{chertkov1,turitsyn,puliafito,celani1}.  The scaling can
be derived by simple physical arguments used by the
theory\cite{chertkov1,celani1}.  The dynamic equation for one of
the molecule orientation angles (see Fig.1A), namely $\phi$, in
the region of $\phi\ll 1$ can be written as
$\frac{d\phi}{dt}=-s\sin^2\phi+\frac{R_g}{R}(\tau_{rel})^{-1/2}\eta_{\phi}
$\cite{chertkov1,turitsyn,puliafito,celani1}, where $R_g$ is the
polymer gyration radius, and $\eta_{\phi}$ is the white thermal
noise. In a stationary state one gets from a balance of shear and
noise terms the following estimates for the rms of the angular
fluctuations (or the PDF width), $\Delta\phi$, and the
characteristic tumbling time, $\tau_t$:

$$\Delta\phi\propto
Wi^{-\frac{1}{3}}(\frac{R_g}{R})^{\frac{2}{3}},\hspace{5 mm}
\tau_t\propto
\tau_{rel}Wi^{-\frac{2}{3}}(\frac{R_g}{R})^{-\frac{2}{3}}.\eqno(1)$$

  Then one can consider two limits at $Wi\gg 1$ (or small $\Delta\phi$ and $\Delta\theta$). In the
linear extension limit, $R\ll R_{max}$, where $R_{max}$ is the
maximum polymer extension, a typical polymer elongation, $R$, can
be estimated as $R_g\cdot Wi$ that is realized at sufficiently
small extensions (up to $20\%$ of $R_{max}$\cite{chu1,sergiy}). It
leads to\cite{chertkov1,turitsyn,puliafito,celani1}

$$\Delta\phi\propto Wi^{-1},\hspace{.5in} \tau_t\propto\tau_{rel}.\eqno(2)$$

In the opposite limit of a non-linear regime, where typically
$Wi\geq R_{max}/R_g$, a stretched polymer can be treated as a
rigid rod with $R\sim R_{max}$ subjected to the Brownian
motion\cite{hinch}. Then the resulting scaling appears to be
strikingly different\cite{chertkov1,turitsyn,puliafito,celani1} :

$$\Delta\phi\propto Wi^{-1/3},\hspace{.5in} \tau_t\propto
\tau_{rel}Wi^{-2/3}. \eqno(3)$$

 Similar scalings can be also expected for $\Delta\theta$.

The experiments were carried out in a shear flow in two flow
configurations. To measure $\theta$ we used a shear flow in a gap
between the flat bottom of the uniformly rotating glass rod of
radius $r_1=1.5$ mm and a cover slip (see for details of the
set-up\cite{sergiy}).  The measurements were made at the fixed
 height of $\sim 30 \mu$m above the cover glass in the region located at  radius $r\simeq 700 \mu$m, where
 the main shear occurs in the vertical plane. Then  only $\theta$ can be measured in this configuration
  (see Fig.1B).

 To measure $\phi$ a shear
 boundary layer of a Poiseuille flow in a micro-channel was
 used. The micro-channel of $\sim 500 \mu$m wide and $\sim 100 \mu$m deep
 was produced from a single cast of a silicon elastomer Polydimethylsiloxane (PDMS) using
 the soft lithography method\cite{xia}. The measurements were carried out at about the middle height of the
 channel in a small region at a distance of $\sim 30 \mu$m  from the channel wall, where only $\phi$
  was detected via the cover glass (see Fig.1C).

Particle image velocimetry (PIV) measurements were conducted with
$0.2 \mu$m fluorescent beads in both flow configurations. In the
swirling flow the shear rate in the vertical plane is rather
constant at the location of measurements of $\sim 30 \mu$m
(Fig.2A), while the azimuthal velocity in the horizontal plane
almost independent of radius in the region of measurements (inset
in Fig.2A). In the channel flow the longitudinal velocity measured
in the horizontal plane changes rather linear in the observation
region of $\sim 30 \mu$m from the wall, while looks fairly
constant in the vertical plane at the middle height of the channel
within the focus depth (0.4 $\mu$m) (see Fig.2B). In both cases
the measured velocity profiles are compared with numerical
calculations, and good quantitative agreement is found (solid
lines, Fig.2A,B). A limitation to conduct PIV measurements in a
vertical plane deeper than $\sim 50 \mu$m in both geometries is
due to the objective working distance that is about $200 \mu$m.

   As a working fluid a buffer solvent\cite{buffer} with sucrose concentration varied
   in the range between $47\%$ (w/w)and $67\%$ (w/w) to tune viscosity in the range between
  $\eta_s=0.012$ Pa$\cdot$s and $\eta_s=0.24$ Pa$\cdot$s at the working temperature of
   22.5$^{\circ}C$ was used.

To study the tumbling dynamics and statistics of polymer molecules
in a flow $10^{-3}$ ppm of $\lambda$-DNA molecules, fluorescently
labelled with YOYO-1 (Molecular Probes) at a dye/base ratio of 1:4
for $\sim 1$ hour, were added into the solvent. At equilibrium the
coiled $\lambda$-DNA has $R_g=0.73 \mu$m, while the entire contour
length is $R_{max}\approx 21 \mu$m. So it may be considered as a
"flexible" polymer with roughly 300 persistence
lengths\cite{chu1}. The maximum relaxation time relevant for
further analysis, $\tau_{rel}=11\pm 0.1$ sec for a solvent with
$62\%$ (w/w) sucrose was found. $\tau_{rel}$ for solvents with
different sucrose concentrations was taken proportional to its
concentration variation\cite{chu1,chu2}. Fluorescently labelled
DNA molecules were monitored via $\times 63$, 1.4NA oil immersion
objective (Zeiss) with 0.4 $\mu$m depth of focus with a homemade
inverted epi-fluorescent microscope\cite{sergiy}. Images of the
molecules were digitized, and their $\theta$ and $\phi$ angles
were automatically measured by approximation of a molecule outline
with an ellipse, which main axis was used as the end-to-end
vector, $\vec{R}$, to define an inclination angle, either $\theta$
or $\phi$ depending on the flow configuration (see Fig.1A,B,C).
This approximation works well for sufficiently stretched molecules
down to $2\div 3 \mu$m for the main axis. Two time series of
different scenarios of tumbling that occur in a shear flow are
presented in Fig.1D. One mentions that tumbling can take a place
via coiled as well as folded states.

In the lower inset in Fig.3 we present PDFs of $\theta$ for two
values of $Wi$. Each plot is based on up to 30,000 points. It is
clear seen that at low $Wi=1.6$ PDF can be fitted by Gaussian
almost entirely, while at high $Wi=17.6$ the PDF tails strongly
deviate from the Gaussian. In Fig.3 PDFs for three values of $Wi$
are presented in the logarithmic coordinates. The PDF tails for
all values of $Wi$ decay algebraically with the exponent rather
close to the theoretical, $|\theta|^{-2}$, indicated by a solid
line on the plot. The larger $Wi$, the wider the scaling region at
the tail as predicted by the theory\cite{chertkov1}. Variations in
$Wi$ values were achieved by viscosity tuning as well as variation
in $s$ by changing rotation speed $\Omega$ between 0 and 0.7
sec$^{-1}$. The half-height width of PDFs, $\Delta\theta$,
decreases
 as $\propto Wi^{-0.38\pm 0.03}$ (see the upper inset in
Fig.3) in a rather good agreement with the recent numerical
simulations based on the FENE model\cite{celani1}.

In the same set-up with the rotating rod we measured PDFs of the
tumbling time $P(\tau)$, determined in the experiment as time
between two following consequent conformations: two coiled, two
folded, or coiled and folded ones (see Fig.1D). This tumbling
time, $\tau$, is elongation based time defined by a certain length
threshold $R_{th}>R_g$, and it is different from $\tau_{\phi}$
that is angular based. The lower the threshold, the higher
$\tau_t$ \cite{celani1}. We present in Fig.4 $P(\tau)$ for three
values of $Wi$ together with the exponential fits. Each plot is
based on up to 600 points. The same data presented in the
log-linear coordinates together with the fits demonstrate the fit
quality (inset in Fig.4). Figure 5 shows the characteristic
tumbling time, $\tau_{t}$, obtained from the slopes of the PDF
tails, versus $Wi$. The value $\tau_t$ depends on the threshold
value $R_{th}/R_g$. $Wi$ was varied in two ways: full squares
present the data, where $Wi$ was tuned just by viscosity variation
(it means the variation of $\tau_{rel}$), and open squares were
obtained at different $Wi$ by adjusting $s$. By presenting the
same data as the ratio $\tau_{rel}/\tau_t$ versus $Wi$ in the
inset in Fig.5 we show, first, that $\tau_t$ is proportional to
$\tau_{rel}$ at lower $Wi$, and second, $\tau_{rel}/\tau_t$ grows
as $\propto Wi^{0.54\pm 0.04}$ at higher $Wi$ in a rather close
agreement with Eqs.(3,4). We would like also to point out the fact
that the cross-over region for scaling in $\tau_t$ takes place at
$Wi\simeq 10$ in amazing agreement between the experiment and the
recent numerical simulations\cite{celani1}. An obvious shift
upwards of the data relatively to the simulations (inset in Fig.5)
is explained by about twice smaller value of $R_{th}/R_g$ for
determination of $\tau$ in the simulations\cite{celani1}.

In Fig.6 PDF of $\phi$ measured in a micro-channel flow at $Wi=25$
together with the fit by $(\sin\phi)^{-2}$ is shown. It is clear
that the PDF maximum is located at $\phi_t\neq 0$ but limited
angular resolution does not allow us to study its dependence on
$Wi$. Similar data on $P(\phi)$, based on 2000-3000 points for
each plot, were obtained for several values of $Wi$.
 The half-height width
of PDFs, $\Delta\phi$, as a function of $Wi$ is shown in the inset
in Fig.6 together with the fit and the results of the numerical
simulations based on the FENE model\cite{celani1}. The observed
scaling $\propto Wi^{-0.51\pm 0.04}$ manifests the cross-over
region between $\propto Wi^{-1/3}$ expected at high $Wi$ and
$Wi^{-1}$ at lower $Wi$ values (see Eq.(3,4)).

 We thank M. Chertkov, V. Lebedev, K. Turitsyn, A. Puliafito, and A. Celani for numerous discussions,
exchange of results of the numerical calculations, and theoretical
guidance, and E. Segre for help with software. This work was
supported by the grants of Minerva Foundation, Israel Science
Foundation, and by the Minerva Center for Nonlinear Physics of
Complex Systems.

\begin{figure}
\caption {A) Polymer orientation angles in a shear flow; B)
Schematic diagram of rotating rod set-up; C) Schematic diagram of
micro-channel set-up; D) Images of various polymer conformations
consequent in time during tumbling. White bar is $3.5 \mu$m.}
 \label{figa}
 \end{figure}

\begin{figure}
\caption {A) Azimuthal velocity vs vertical coordinate; inset: the
same vs radius. Region of measurements is around $R=700 \mu$m and
at height of $\sim 30 \mu$m. B) Squares are longitudinal velocity
in horizontal plane, circles- the same in vertical plane. Region
of measurements at $30\pm 10 \mu$m from the channel wall and at
height of $\sim 50 \mu$m (dash line). Solid lines on all plots are
theoretical.}
 \label{figb}
 \end{figure}

\begin{figure}
\caption {PDFs of $|\theta|$ in logarithmic coordinates:
$Wi=1.6$-pluses, $Wi=8.5$-circles, $Wi=17.6$-squares. A solid line
is theoretically predicted asymptotic scaling for PDF tails. Lower
inset: PDFs of $\theta$ for $Wi=1.6$-pluses, and $Wi=17.6$-squares
presented in log-linear coordinates; the solid curves are Gaussian
fits. Upper inset: $\Delta\theta$ vs $Wi$; squares-data,
triangles-numerics. A solid line is the fit.}

\label{figc}
 \end{figure}

\begin{figure}
\caption {PDFs of $\tau$ at several values of $Wi$. The solid
lines are exponential fits. Inset: PDFs of $\tau$ in log-linear
coordinates with fits to the data at $Wi=4.7$-triangles,
$Wi=8.5$-circles, $Wi=17.6$-squares.}
 \label{figd}
 \end{figure}

 \begin{figure}
\caption { $\tau_t$ vs $Wi$: full squares-data taken at constant
$s$ and varying viscosity; open squares-data taken at constant
viscosity and varying $s$. The dash lines are guides to eye.
Inset: $\tau_{rel}/\tau_t$ vs $Wi$; triangles-numerics. Solid
lines are the fits.}
 \label{fige}
 \end{figure}

 \begin{figure}
\caption {PDFs of $\phi$ in log-linear coordinates at $Wi=25$. A
solid curve is the fit by $\propto (\sin\phi)^{-2}$. Upper inset:
$\Delta\phi$ vs $Wi$; squares-data, triangles-numerics. A solid
line is the fit.}
 \label{figef}
 \end{figure}

 \end{multicols}

\begin{references}
\bibitem{buffer}{A pH8 buffer
 consisting of 10mM tris-HCl, 2 mM EDTA, 10 mM NaCl, $4\% \beta$-mercaptoethanol, glucose
oxidase ($\sim 50 \mu g/ml$) and catalase ($\sim 10 \mu g/ml$),
 $62\%$ (w/w) sucrose was used (see e.g. Ref.\cite{chu1}).}
\end{references}
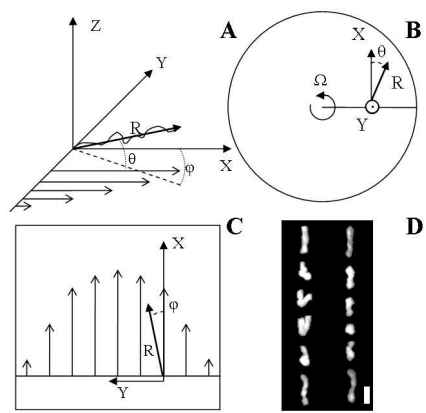
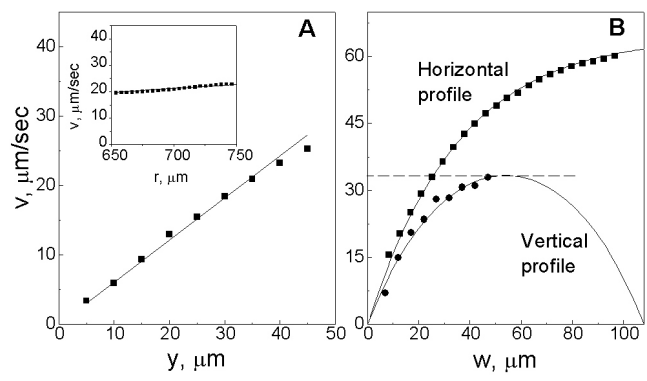
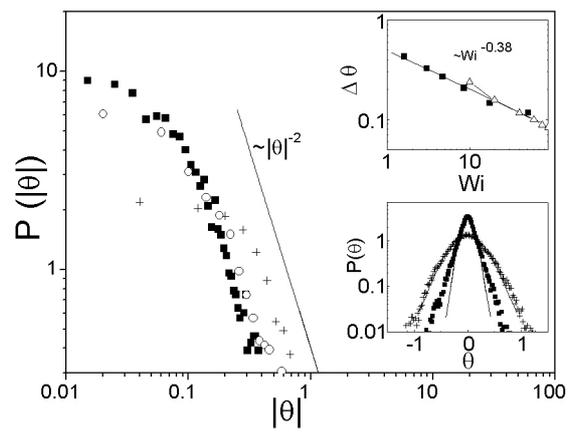
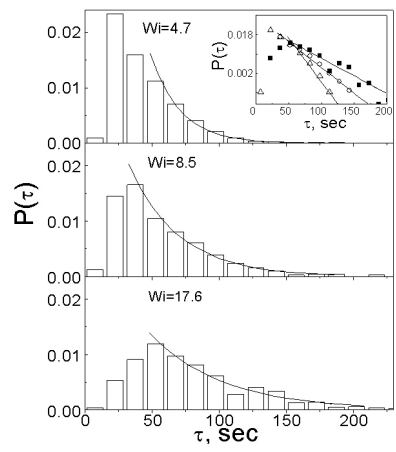
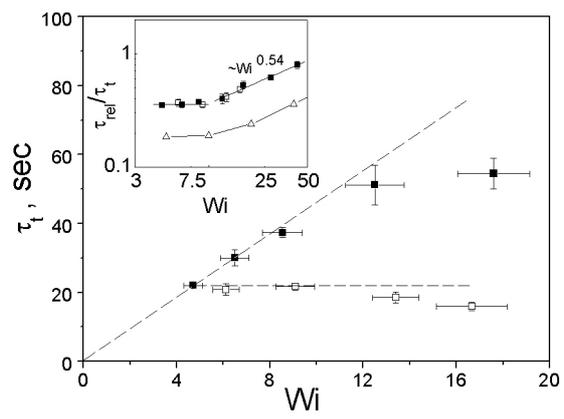
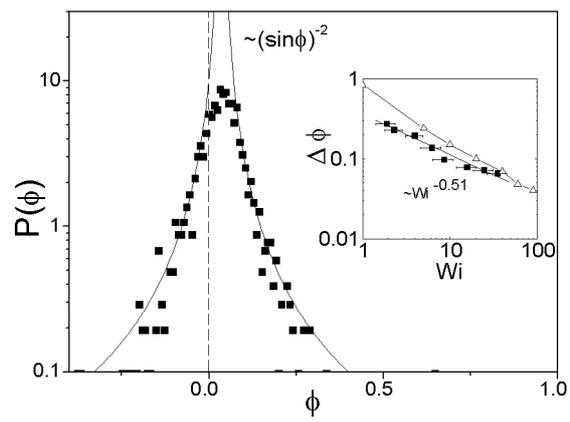
\end{document}